\documentclass[aps,prd,preprint,12pt,superscriptaddress,nofootinbib,floatfix]{revtex4}  
\pdfoutput=1
\usepackage{fancyhdr}
\usepackage{latexsym}
\usepackage{amsfonts}

\usepackage[latin1]{inputenc}
\usepackage{graphicx}
\usepackage{mathrsfs}
\usepackage{amssymb}
\usepackage{amsmath}
\usepackage{verbatim}
\usepackage{multirow}
\usepackage{soul}
\usepackage{subfigure}
\usepackage[usenames,dvipsnames]{color}

\newcommand{\TeV}{{\ensuremath\rm TeV}}
\newcommand{\GeV}{{\ensuremath\rm GeV}}

\newcommand{\fb}{{\ensuremath\rm fb}}
\newcommand{\pb}{{\ensuremath\rm pb}}
\newcommand{\eqn}{equation}
\newcommand{\al}{\alpha}
\newcommand{\be}{\beta}
\newcommand{\lb}{\left(}
\newcommand{\rb}{\right)}

\newcommand{\lam}{\lambda}

\def\D0{\slash\!\!\!\!\!\!\!\!\!\:D0}
\newcommand{\HS}{\texttt{HiggsSignals}}
\newcommand{\HB}{\texttt{HiggsBounds}}
\newcommand{\HSv}[1]{\texttt{HiggsSignals-#1}}

\newcommand{\cp}{\mathcal{CP}}

\newcommand{\Htohh}{H\to hh}

\newcommand{\oblique}{Altarelli:1990zd,Peskin:1990zt,Peskin:1991sw,Maksymyk:1993zm}

\begin{document}

\date{\today}
\title{{\Large LHC Benchmark Scenarios for the\\[0.2cm]
Real Higgs Singlet Extension of the Standard Model}\\[0.5cm]
}

\author{Tania Robens}
\email{Tania.Robens@tu-dresden.de}
\affiliation{TU Dresden, Institut f\"ur Kern- und Teilchenphysik,
Zellescher Weg 19, D-01069 Dresden, Germany\vspace{0.2cm}}
\author{Tim Stefaniak\vspace{0.2cm}}
\email{tistefan@ucsc.edu}
\affiliation{Department of Physics and Santa Cruz Institute for Particle Physics, University of California, Santa Cruz, CA 95064, USA\vspace{0.5cm}}
\renewcommand{\abstractname}{\vspace{0.5cm} Abstract}

\begin{abstract}
\vspace{0.5cm}
We present benchmark scenarios for searches for an additional Higgs state in the real Higgs singlet extension of the Standard Model in Run 2 of the LHC. The scenarios are selected such that they fulfill all relevant current theoretical and experimental constraints, but can potentially be discovered at the current LHC run. We take into account the results presented in earlier work and update the experimental constraints from relevant LHC Higgs searches {and signal rate measurements}. {The} benchmark scenarios are given separately for the low mass and high mass region, i.e.~the mass range where the additional Higgs state is lighter or heavier than the discovered Higgs state at around $125$ GeV. The{y} have also been presented in the framework of the LHC Higgs Cross Section Working Group.
\end{abstract}
\preprint{SCIPP 16/03}

\maketitle

\section{Introduction}
\label{Sec:Intro}
\noindent
{The first run of the LHC at center-of-mass (CM) energies of $7$ and $8\,\TeV$ has been completed in 2015. Its remarkable success is highlighted by the breakthrough discovery of a scalar boson in July 2012 and the measurements of its coupling properties, which thus far} are well compatible with the interpretation in terms of the Higgs boson of the Standard Model (SM) Higgs mechanism~\cite{Higgs:1964ia,Higgs:1964pj,Englert:1964et, Guralnik:1964eu, Kibble:1967sv}. {The} combination of the Higgs mass measurements performed by ATLAS and CMS yields \cite{Aad:2015zhl}
\begin{align}
m_H = 125.09\,\pm\,0.21\,\text{(stat.)}\pm\,0.11\,\text{(syst.)}~\GeV.
\label{Eq:mhexp}
\end{align}
If the discovered particle is indeed the Higgs boson {of} the SM, {its mass measurement determines} the last unknown ingredient {of} this model, as all other properties of the electroweak sector then follow directly from theory. {In the coming years a thorough investigation} of the Higgs boson's properties {is needed} in order to identify whether the SM Higgs sector is indeed complete, or instead, the structure of a more involved Higgs sector is realized. {This includes} detailed and accurate measurements of its {coupling strengths and $\cp$ structure} at the LHC and {ultimately} at future experimental facilities for Higgs boson precision {studies}. {Complementary to this,} collider searches for additional Higgs bosons {need to be continued over the full accessible mass range}. {The discovery of another Higgs boson} would {inevitably prove the existence of} a non-minimal Higgs sector. \\
In this work we consider the simplest extension of the SM Higgs sector, where an additional real {scalar} field is added, which is neutral under all quantum numbers of the SM gauge groups~\cite{Schabinger:2005ei,Patt:2006fw} and acquires a vacuum expectation value (VEV). This model has been widely studied in the literature~\cite{Barger:2007im, Bhattacharyya:2007pb, Dawson:2009yx, Bock:2010nz,Fox:2011qc, Englert:2011yb,Englert:2011us,Batell:2011pz, Englert:2011aa, Gupta:2011gd, Dolan:2012ac, Bertolini:2012gu,Batell:2012mj,Lopez-Val:2013yba,Heinemeyer:2013tqa,Chivukula:2013xka,Englert:2013tya,Cooper:2013kia,Caillol:2013gqa,Coimbra:2013qq,Pruna:2013bma,Dawson:2013bba, Basso:2013nza,Lopez-Val:2014jva,Englert:2014aca,Englert:2014ffa,Chen:2014ask,Karabacak:2014nca,Profumo:2014opa,Robens:2015gla,Martin-Lozano:2015dja,Falkowski:2015iwa,Ballesteros:2015iua,Buttazzo:2015bka,Banerjee:2015hoa,Corbett:2015lfa,Tofighi:2015fia,Chen:2015gva,Godunov:2015nea,Duch:2015jta,Wang:2015cda,Bernal:2015xba,Ghosh:2015apa,Dolan:2016eki}, also in the context of electroweak higher order corrections \cite{Kanemura:2015fra,Bojarski:2015kra} or offshell and interference effects \cite{Englert:2014aca,Englert:2014ffa,Maina:2015ela,Kauer:2015hia,Englert:2015zra,Ballestrero:2015jca,Dawson:2015haa}. Here, we present an update of the exploration of the model parameter space presented in Ref.~\cite{Robens:2015gla}, where we take the latest experimental constraints 
into account. As before, we consider masses of the second (non-standard) Higgs boson in the whole mass range up to $1\,\TeV$. This minimal setup can be interpreted as a limiting case for more generic BSM scenarios, e.g.~models with additional gauge sectors~\cite{Basso:2010jm} or additional matter content~\cite{Strassler:2006im,Strassler:2006ri}. Experimental searches for the model have been presented in \cite{ATLAS:2014rxa,ATLAS:2014kua,Aad:2014ioa,Khachatryan:2015cwa,Aad:2015xja,Aad:2015pla,ATLAS-CONF-2015-081,Aad:2015kna}.
  
{As in} Ref.~\cite{Robens:2015gla} {we take} the following theoretical and experimental constraints into account: bounds from perturbative unitarity and electroweak (EW) precision measurements, {in particular} focus{sing} on higher order corrections to the $W$ boson mass~\cite{Lopez-Val:2014jva}; perturbativity, vacuum stability and correct minimization of the model up to a high energy scale using renormalization group (RG) evolved couplings; exclusion limits from Higgs searches at the LEP, Tevatron and LHC experiments via the public tool \HB~\cite{Bechtle:2008jh,Bechtle:2011sb,Bechtle:2013gu,Bechtle:2013wla,Bechtle:2015pma}, and  compatibility of the model with the signal strength measurements of the discovered Higgs state using \HS~\cite{Bechtle:2013xfa} (cf.~also Ref.~\cite{Bechtle:2014ewa}).
  
We separate the discussion of the parameter space into two different mass regions: \emph{(i)} the high mass region, $m_H \in [130, 1000]\,\GeV$, where the lighter Higgs boson $h$ is interpreted as the discovered Higgs state;  \emph{(ii)} the low mass region, $m_h \in [1,120]\,\GeV$, where the heavier Higgs boson $H$ is interpreted as the discovered Higgs state.
  
 We find that the most severe constraints in the whole parameter space for the second {Higgs} mass {$m_H \lesssim 250~\GeV$} are {mostly} given by limits from {collider searches for a SM Higgs boson} as well as {by} the {LHC} Higgs {boson} signal strength {measurements}. For $m_H\gtrsim 250\,\GeV$ limits from higher order contributions to the $W$ boson mass prevail, followed by the requirement of perturbativity of the couplings. 
 
 {For the remaining viable parameter space} we present predictions for {signal} cross sections of the {yet undiscovered second} Higgs boson {for the LHC at a CM energy of} $14\,\TeV${, discussing both the SM Higgs decay signatures and the novel Higgs-to-Higgs decay mode $\Htohh$.} For both the high mass and low mass region we present a variety of {\sl benchmark scenarios}. These are designed to render a maximal direct production rate for the {collider signature of interest}. {Whenever kinematically accessible we give} two different benchmark points for each mass, {for which} the {Higgs-to-Higgs} decay $\Htohh$ is maximal or minimal, respectively.

The paper is organized as follows: In Section~\ref{Sec:Model} we briefly review the model {and the chosen parametrization}. {In} Section~\ref{sec:constraints} we review the constraints {that are taken into account} and in particular discuss the impact of the new constraints on the parameter space. In Section~\ref{sec:bms} we provide benchmark points and planes discussed above. We summarize {and conclude} in Section~\ref{sec:conclude}.


\section{The model}
\label{Sec:Model}
In the following we briefly review the main features of the real Higgs singlet extension of the SM that are important for the benchmark choices. More details about the model can e.g.~be found in Refs.~\cite{Pruna:2013bma,Lopez-Val:2014jva,Robens:2015gla,Bojarski:2015kra} and references therein.
\subsection{Potential and couplings}
The real Higgs singlet extension of the SM \cite{Schabinger:2005ei,Patt:2006fw, Bowen:2007ia} contains a complex $SU(2)_L$ doublet, in the following denoted by $\Phi$, and in additional a real scalar $S$ which is a singlet under the SM gauge group. The most general renormalizable Lagrangian compatible with an additional $Z_2$ symmetry is then given by
\begin{equation}\label{lag:s}
\mathscr{L}_s = \left( D^{\mu} \Phi \right) ^{\dagger} D_{\mu} \Phi + 
\partial^{\mu} S \partial_{\mu} S - V(\Phi,S ) \, ,
\end{equation}
with the scalar potential
\begin{eqnarray}\label{potential}\nonumber
V(\Phi,S ) &=& -m^2 \Phi^{\dagger} \Phi - \mu ^2  S ^2 +
\left(
\begin{array}{cc}
\Phi^{\dagger} \Phi &  S ^2
\end{array}
\right)
\left(
\begin{array}{cc}
\lambda_1 & \frac{\lambda_3}{2} \\
\frac{\lambda_3}{2} & \lambda _2 \\
\end{array}
\right)
\left(
\begin{array}{c}
\Phi^{\dagger} \Phi \\  S^2 \\
\end{array}
\right) \\
\nonumber \\ 
&=& -m^2 \Phi^{\dagger} \Phi -\mu ^2 S ^2 + \lambda_1
(\Phi^{\dagger} \Phi)^2 + \lambda_2  S^4 + \lambda_3 \Phi^{\dagger}
\Phi S ^2.
\end{eqnarray}
{The implicitly imposed $Z_2$ symmetry} forbids all linear or cubic terms of the singlet field $S$ in the potential. We assume that both Higgs fields {$\Phi$} and $S$ have a non-zero vacuum expectation value (VEV), {denoted by $v$ and $x$, respectively.}
In the unitary gauge, the Higgs fields are given by
\begin{equation}\label{unit_higgs}
\Phi \equiv
\left(
\begin{gathered}
0 \\
\tfrac{\tilde{h}+v}{\sqrt{2}}
\end{gathered} \right), 
\hspace{2cm}
S \equiv \frac{h'+x}{\sqrt{2}}.
\end{equation} 
After diagonalization of the mass matrix we obtain the mass eigenstates $h$ and $H$ with mass eigenvalues given by
\begin{align}\label{mh1}
m^2_{h} &= \lambda_1 v^2 + \lambda_2 x^2 - \sqrt{(\lambda_1 v^2 -
  \lambda_2 x^2)^2 + (\lambda_3 x v)^2}, \\
\label{mh2}
m^2_{H} &= \lambda_1 v^2 + \lambda_2 x^2 + \sqrt{(\lambda_1 v^2 -
  \lambda_2 x^2)^2 + (\lambda_3 x v)^2},
\end{align}
and $m^2_{h} \le m^2_{H}$ by convention. The gauge and mass eigenstates are related via the mixing matrix
\begin{equation}\label{eigenstates}
\left(
\begin{array}{c}
h \\
H
\end{array}
\right) = \left(
\begin{array}{cc}
\cos{\alpha} & -\sin{\alpha} \\
\sin{\alpha} & \cos{\alpha}
\end{array}
\right) \left(
\begin{array}{c}
\tilde{h} \\
h'
\end{array}
\right),
\end{equation}
where {the mixing angle} $-\frac{\pi}{2} \leq \alpha \leq \frac{\pi}{2}$ {is given by}
\begin{align}\label{sin2a}
\sin{2\alpha} &= \frac{\lambda_3 x v}{\sqrt{(\lambda_1 v^2 -
    \lambda_2 x^2)^2 + (\lambda_3 x v)^2}}, \\
\label{cos2a}
\cos{2\alpha} &= \frac{\lambda_2 x^2 - \lambda_1
  v^2}{\sqrt{(\lambda_1 v^2 - \lambda_2 x^2)^2 + (\lambda_3 x v)^2}}.
\end{align}
It follows from Eq.~\eqref{eigenstates} that the light (heavy) Higgs boson couplings to SM particles are suppressed by $\cos\al\,(\sin\al)$.

If kinematically allowed, the additional decay channel $\Htohh$ is present. Its {partial} decay width at leading order (LO) is given by~\cite{Schabinger:2005ei, Bowen:2007ia}
\begin{align}\label{eq:gtot}
\Gamma_{H\rightarrow hh}\,=\,\frac{|\mu'|^2}{8\pi m_{H}}\,\sqrt{1-\frac{4 m^2_{h}}{m_{H}^2}} \, ,
\end{align}
where the coupling strength $\mu'$ of the $H \rightarrow hh$ decay reads
 \begin{align}
\mu'=-\frac{\sin\lb 2\al \rb}{2vx}\,\lb \sin\al v+ \cos\al\,x\rb\,\lb m_h^2+\frac{m_H^2}{2} \rb.
\label{eq:muprime}
\end{align}
Next-to-leading order (NLO) corrections to the $\Htohh$ decay width for this model have been calculated recently in Ref.~\cite{Bojarski:2015kra}.
The branching ratios of the {\sl heavy} Higgs mass eigenstate $m_H$ are then given by
\begin{eqnarray}
\text{BR}_{H\rightarrow hh}&=&\frac{\Gamma_{H\rightarrow hh}}{\Gamma_\text{tot}}, \\
\text{BR}_{H\rightarrow \text{SM}}&=&\sin^2\al \times \frac{\Gamma_{\text{SM}, H\rightarrow\text{SM}}}{\Gamma_\text{tot}},\label{eq:brdefs}
\end{eqnarray}
where $\Gamma_{\text{SM},\,H\rightarrow\text{SM}}$ is the partial decay width of the SM Higgs boson {and $H\to\text{SM}$ represents any SM Higgs decay mode}. The total width is then
\begin{\eqn}
\Gamma_\text{tot}\,=\,\sin^2\al\,\times\,\Gamma_\text{SM, tot}+\Gamma_{\Htohh},
\end{\eqn}
where $\Gamma_\text{SM, tot}$ denotes the total width of the SM Higgs boson with mass $m_H$. The suppression by $\sin^2\,\al$ directly follows from the suppression of all SM--like couplings, cf.~Eq.~\eqref{eigenstates}. For $\mu'\,=\,0$, the decay $\Htohh$ vanishes and we recover the SM Higgs boson branching ratios.

For the collider phenomenology of the model two features are important:

\begin{itemize}
\item the suppression of the {\sl production {cross section}} of the
  two {Higgs states induced by the mixing}, which {is} given by $\sin^2\al\,(\cos^2\al)$ for the heavy (light) Higgs, respectively;
\item {the} suppression of the {\sl {Higgs} decay modes to SM particles}, which is {realized} if the competing decay mode $\Htohh$ is kinematically accessible. 

\end{itemize}
For the high mass (low mass) scenario, {i.e.~the case where the light (heavy) Higgs boson is identified with the discovered Higgs state at $\sim 125~\GeV$}, $|\sin\al|\,=\,0\,(1)$ corresponds to the complete decoupling of the second Higgs {boson} and therefore the SM-like scenario.

\subsection{Model parameters}
\noindent
At the Lagrangian level, the model has five free parameters,
\begin{\eqn}
\lambda_1,\,\lambda_2,\,\lambda_3,\,v,\,x,
\end{\eqn}
while the values of the additional parameters $\mu^2,\,m^2$ are fixed by the minimization conditions. 
A more intuitive basis, where the free model parameters are represented by physical (i.e.~observable) quantities, is given by\footnote{Note that even if the $Z_2$ symmetry is not imposed, the parameters of the model relevant for the collider phenomenology considered here can always be chosen {in terms of} the masses, a mixing angle, and an additional parameter determining the $\Htohh$ decay channel.}
\begin{\eqn}\label{eq:pars}
m_h,\,m_H,\,\sin\alpha,\,v,\,\tan\beta\,\equiv\,\frac{v}{x}.
\end{\eqn}
The vacuum expectation value of the Higgs doublet {$\Phi$} is given by the SM value $v~\sim~246~\GeV$, and one of the Higgs masses is fixed to {$m_{h/H}\,=\,125.09\,\GeV$}, eliminating two of the five parameters. We are thus left with only three independent parameters, 
\begin{align}
\left\{ {m\equiv m_{H/h}},\,\sin\alpha,\,\tan\be \right\}, \label{eq:par_choices1}
\end{align}
where the latter enters the collider phenomenology only through the heavy Higgs decay mode into the lighter Higgs, $\Htohh$. Note that from a collider perspective, for cases where the decay mode $\Htohh$ is kinematically allowed, the input parameter $\tan\beta$ could be replaced by either the total width of the heavier state, $\Gamma(H)$, the branching ratio $\text{BR} \lb \Htohh \rb$, or the partial decay width of this channel, $\Gamma(\Htohh)$, respectively, rendering the following viable parameter choices besides Eq.~\eqref{eq:par_choices1}:
\begin{align}
&\left\{m\equiv m_{H/h},\,\sin\alpha,\,\Gamma(H)\right\},\label{eq:par_choices2}\\
&\left\{m\equiv m_{H/h},\,\sin\alpha,\,\text{BR} (\Htohh)\right\},\label{eq:par_choices3}\\
&\left\{m\equiv m_{H/h},\,\sin\alpha,\,\Gamma(\Htohh)\right\}.\label{eq:par_choices4}
\end{align}
If the insertion starts on the Lagrangian level (via e.g.~\texttt{FeynRules}~\cite{Christensen:2008py}, \texttt{SARAH}~\cite{Staub:2008uz,Staub:2013tta} or similar), also the Lagrangian parameters as such can be used as input values, but then care must be taken to correctly translate these into the phenomenologically viable parameter regions.

\section{Constraints}
\label{sec:constraints}

{In this section we list all theoretical and experimental constraints that we take into account, and give an overview over the impact of these constraints on the parameter space. We refer the reader to Ref.~\cite{Robens:2015gla} for details on the implementation of these constraints. With respect to Ref.~\cite{Robens:2015gla} we update the experimental limits from LHC Higgs searches, leading to a change in the allowed parameter space especially in the lower mass range, $m_H\,\in\,[130, 250 ] \,\GeV$. We also include constraints from the combined ATLAS and CMS Higgs signal strength~\cite{ATLAS-CONF-2015-044}, rendering a significantly stronger limit on the mixing angle. However, this limit is still not as strong as the constraint} from the $W$ boson mass measurement in most of the parameter space.
\subsection{Theoretical Constraints}
We consider the following theoretical constraints in the selection of the benchmark scenarios:
\begin{itemize}
\item{}vacuum stability and minimization of model up to a scale $\mu_\text{run}\,=\,4\,\times\,10^{10}\,\GeV$,
\item{}perturbative unitarity of the $2\to 2$ $S$-matrix for $(W^+\,W^-,ZZ,hh,hH,HH)$ initial and final states,
\item{}perturbativity of the couplings in the potential, $|\lam_i|\,\leq\,4\,\pi$, up to a high energy scale, $\mu_\text{run}\,=\,4\,\times\,10^{10}\,\GeV$, {employing one-loop renormalization group equations (RGEs)~\cite{Lerner:2009xg}.}
\end{itemize}
\subsection{Experimental Constraints}
The following experimental constraints are taken into account at the $95\%$ C.L.:
\begin{itemize}
\item{}agreement with electroweak precision observables, employing the oblique parameters $S,\,T,\,U$ \cite{\oblique} and using the results from the global fit from the \texttt{GFitter} Group~\cite{Baak:2014ora},
\item{}agreement with the observed $W$ boson mass~\cite{Alcaraz:2006mx, Aaltonen:2012bp, D0:2013jba}, $M_W = 80.385 \pm 0.015~\GeV$, employing the NLO calculation presented in Ref.~\cite{Lopez-Val:2014jva},
\item{}agreement with limits from direct Higgs searches at LEP, Tevatron, and the LHC using \HB\ (version 4.3.1)~\cite{Bechtle:2008jh,Bechtle:2011sb,Bechtle:2013gu,Bechtle:2013wla,Bechtle:2015pma}. {With respect to the results presented in Ref.~\cite{Robens:2015gla}, limits from the following searches have been included here:
\begin{itemize}
\item {ATLAS search for $H\to WW$~\cite{Aad:2015agg},}
\item {ATLAS search for $H\to ZZ$~\cite{Aad:2015kna},}
\item combination of ATLAS searches for $\Htohh \to bb\tau\tau, \gamma\gamma WW^*, \gamma\gamma bb, bbbb$~\cite{Aad:2015xja},
\item CMS search for $H\to VV~(V=W^\pm,Z)$~\cite{Khachatryan:2015cwa},
\item {CMS search for $\Htohh\to 4\tau$, where $H$ is the SM-like Higgs boson at $125\,\GeV$~\cite{CMS:2015iga}.}
\end{itemize}
}
\item{}Agreement with the observed signal strengths of the 125 \GeV~Higgs boson, using \HS\ (version 1.4.0)~\cite{Bechtle:2013xfa}, and using the results from the ATLAS and CMS combination of the LHC Run 1 data, {$\mu\,=\, 1.09 \pm 0.11$~\cite{ATLAS-CONF-2015-044}, leading to
\begin{align}
|\sin\al|\,\leq\,0.36\,
\end{align}
for the heavy Higgs mass range $m_H \gtrsim 150\,\GeV$ (high mass range, $m_h\,\sim\,125\,\GeV$), and
\begin{align}
|\sin\al|\,\geq\,0.87\,
\end{align}
for the light Higgs mass range $m_h \lesssim 100\,\GeV$ (low-mass range, $m_H\,\sim\,125\,\GeV $). In these mass regions potential signal overlap with the SM-like Higgs at $125\,\GeV$ can be neglected. For Higgs masses in the range $[100, 150]\,\GeV$ we employ \HS\ using observables from the individual Higgs channels, which enables to approximately take into account a potential signal overlap~\cite{Bechtle:2013xfa}, see also Ref.~\cite{Robens:2015gla} for details.
}
\end{itemize}
\clearpage

\subsection[Parameter Regions and Constraints]{Allowed Parameter Regions and Sensitivity of the Constraints}
\subsubsection{High mass region}
\begin{figure}[b]
\includegraphics[width=\textwidth]{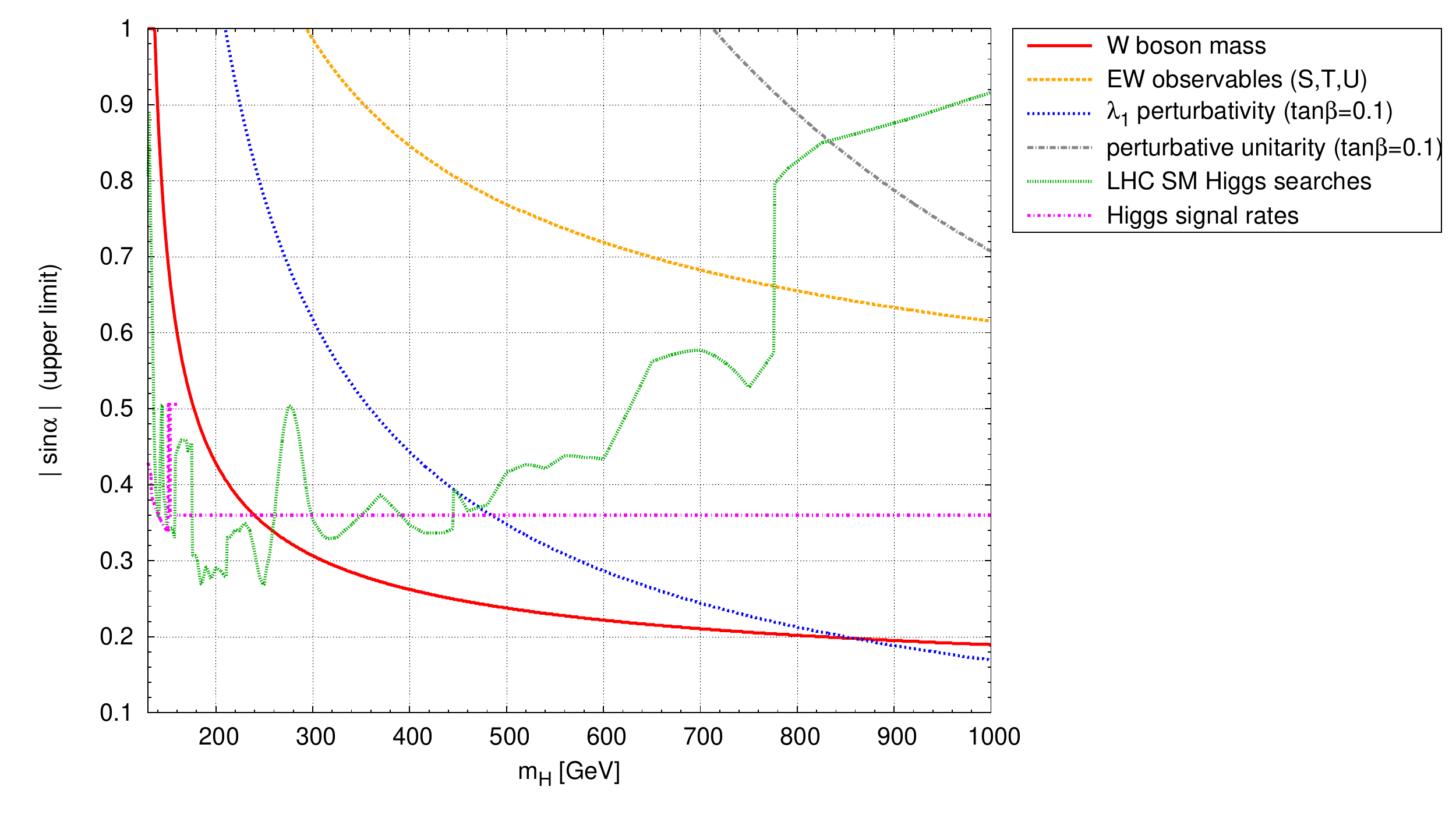}
\caption{\label{fig:sinamw} Maximal allowed values for $| \sin\al |$ in the high mass region, $m_H\in [130, 1000]\,\GeV$, from  {NLO} calculations of the $W$ boson mass (\emph{red, solid})~\cite{Lopez-Val:2014jva}, electroweak precision observables (EWPOs) tested via the oblique parameters $S$, $T$ and $U$ (\emph{orange, dashed}), perturbativity of the RG-evolved coupling $\lam_1$ (\emph{blue, dotted}), evaluated {for an exemplary choice} $\tan\be\,=\,0.1$, perturbative unitarity (\emph{grey, dash-dotted}), direct LHC Higgs searches (\emph{green, dashed}), and the Higgs signal strength (\emph{magenta, dash-dotted}). }
\end{figure}

The importance of the different constraints on the mixing angle $\sin\al$ in the high mass region, where $m_h\,\sim\,125\,\GeV$, is summarized in Figure \ref{fig:sinamw}. {Recall} that this angle {is responsible for} the {\sl global} suppression of the production cross section with respect to {the SM prediction} {at} the same {Higgs} mass. We see that {in the lower mass region, $m_H\,\lesssim\,250\,\GeV$,} the {most} important constraints stem from direct Higgs searches~\cite{Chatrchyan:2013mxa,Khachatryan:2015cwa,CMS:aya,CMS:bxa,Aad:2015kna} and the combined Higgs signal strength~\cite{ATLAS-CONF-2015-044}, {whereas for higher masses, $m_H\,\in\,[250\,\GeV;\;800\,\GeV]$, the $W$ boson mass becomes the strongest constraint}~\cite{Lopez-Val:2014jva}. Requiring perturbativity of the couplings {yields the upper limit on $|\sin\alpha|$ for very heavy Higgs bosons,} $m_H\,\geq\,800\,\GeV$.

\begin{figure}
\includegraphics[width=0.75\textwidth]{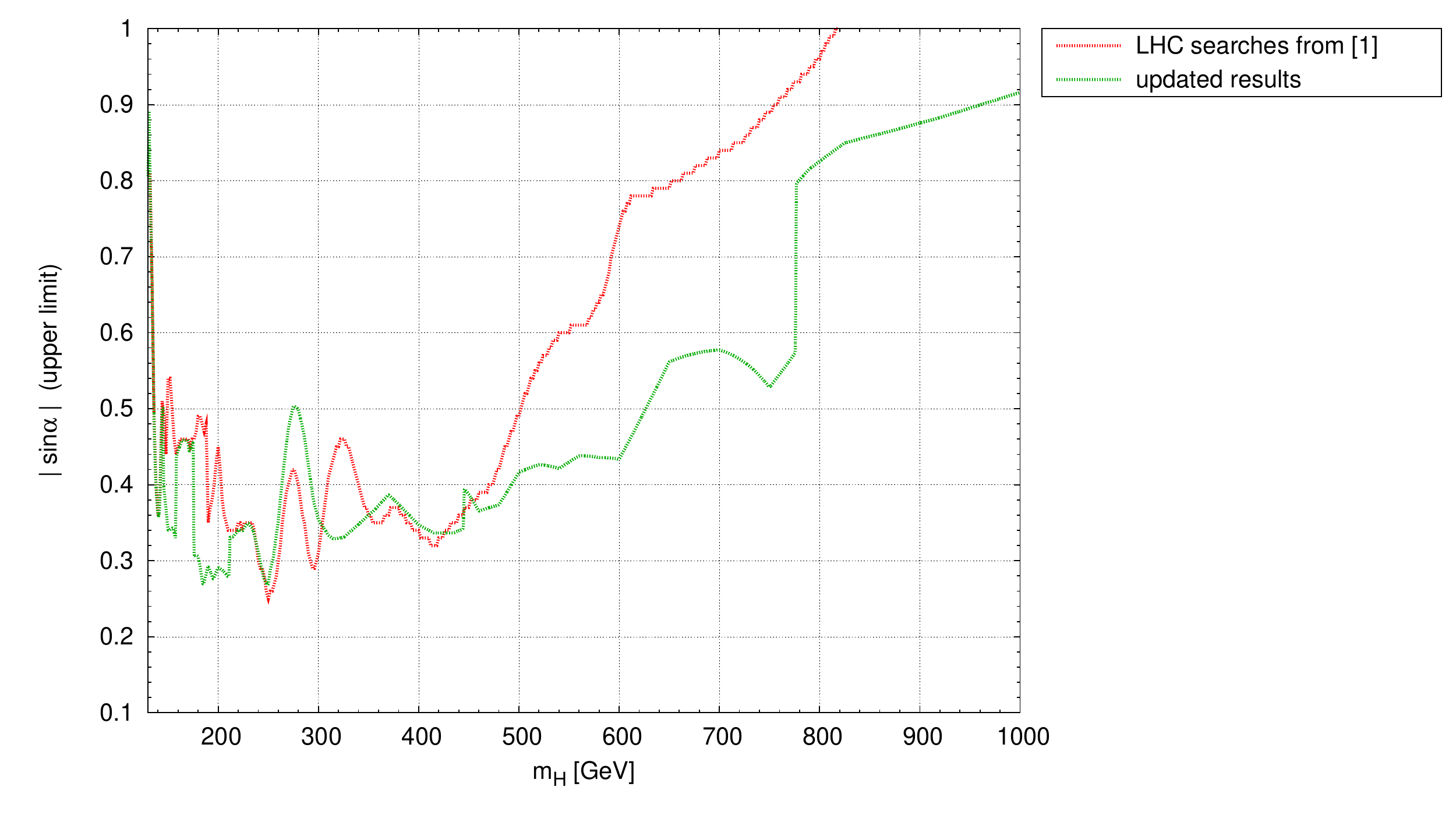}
\caption{\label{fig:comp} {Comparison of the $|\sin\alpha|$ limit obtained from the LHC Higgs searches with SM final states as presented in Ref.~\cite{Robens:2015gla} {\sl (red)} with the updated analysis} {\sl (green)}.}
\end{figure}

The updated combined signal strength reduces the maximally allowed mixing angle from previously $|\sin\al|\,\lesssim\,0.50$~\cite{Robens:2015gla} to $|\sin\al|\,\lesssim\,0.36$. The updated limits from LHC Higgs searches {in channels with vector boson final states} also generally lead to stronger constraints{, except in the region $m_H \in [260,300]\,\GeV$, where a statistical upward fluctuation in the CMS $H\to ZZ\to 4\ell$ channel~\cite{Khachatryan:2015cwa} leads to a slightly weaker limit than previously observed.} A comparison of previously presented {limits from LHC Higgs searches with the current status} is displayed in Fig.~\ref{fig:comp}. {We see that the updated constraints yield stronger limits in particular for $m_H\,\leq\,250\,\GeV$ as well as for $m_H\,\gtrsim\,400\,\GeV$.} {We supplement this comparison by giving a detailed list in Tab.~\ref{tab:channels} of the LHC Higgs search channels that have been applied by \HB\ in the various mass regions.\footnote{{\HB\ selects the most sensitive channel by comparing the expected exclusion limits first. In a second step, the predicted signal strength is confronted with the observed exclusion limit only of this selected channel. This well defined statistical procedure allows to systematically test the model against a plethora of Higgs search limits without diluting the $95\%$ C.L.~of the individual limits.}}}

\begin{table}
\begin{tabular}{|c|c|c|}
\toprule
Range of $m_H\,[GeV]$&Search channel& Reference\\ 
\colrule
130-145 &        H$\rightarrow$ZZ$\rightarrow$4l  & \cite{Chatrchyan:2013mxa} (CMS) \\
145-158  &      H$\rightarrow$VV (V=W,Z)& \cite{Khachatryan:2015cwa} (CMS)\\
158-163  &      SM comb.     &   \cite{CMS:aya} (CMS)\\
163-170  &      H$\rightarrow$WW        &  \cite{CMS:bxa} (CMS)\\
170-176  &      SM comb.     &  \cite{CMS:aya} (CMS)\\
176-211  &      H$\rightarrow$VV (V=W,Z)&  \cite{Khachatryan:2015cwa} (CMS)\\
211-225  &      H$\rightarrow$ZZ$\rightarrow$4l    &  \cite{Chatrchyan:2013mxa} (CMS)\\
225-445  &      H$\rightarrow$VV (V=W,Z)&   \cite{Khachatryan:2015cwa} (CMS)\\
445-776  &      H$\rightarrow$ZZ         &    \cite{Aad:2015kna} (ATLAS)\\
776-1000 &              H$\rightarrow$VV (V=W,Z)&   \cite{Khachatryan:2015cwa} (CMS)\\
\botrule
\end{tabular}
\caption{\label{tab:channels} List of LHC Higgs search channels that are applied by \HB\ in the high mass region, yielding the upper limit on $|\sin\alpha|$ shown in Figs.~\ref{fig:sinamw} and \ref{fig:comp}.}
\end{table}

\begin{figure}
\centering
\subfigure[~{Heavy Higgs signal rate with SM particles in the final state for the LHC at $14~\TeV$}.]{
\includegraphics[width=0.48\textwidth]{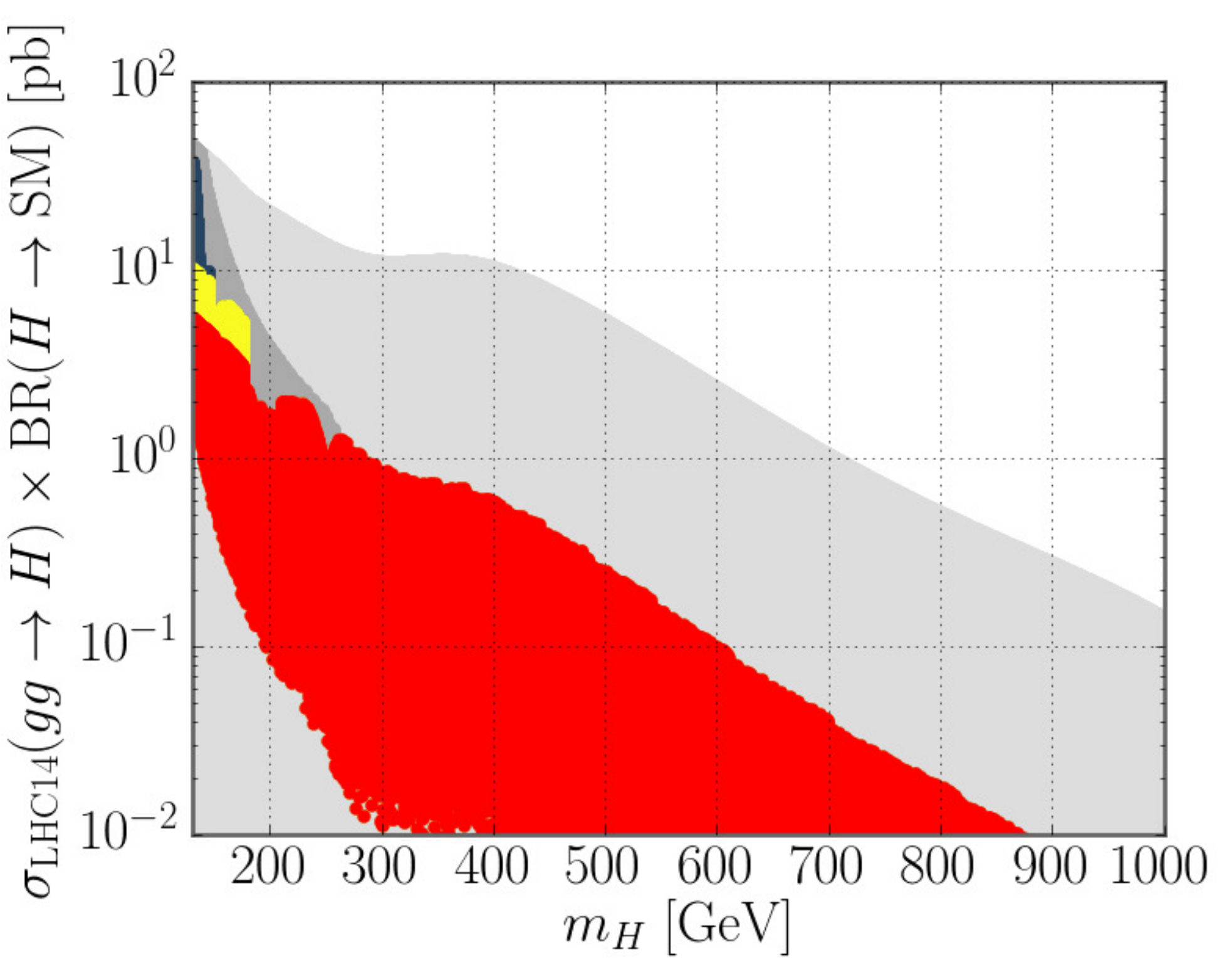}
}
\hfill
\subfigure[~{Heavy Higgs signal rate with light Higgs bosons in the final state for the LHC at $14~\TeV$}.]{
\includegraphics[width=0.48\textwidth]{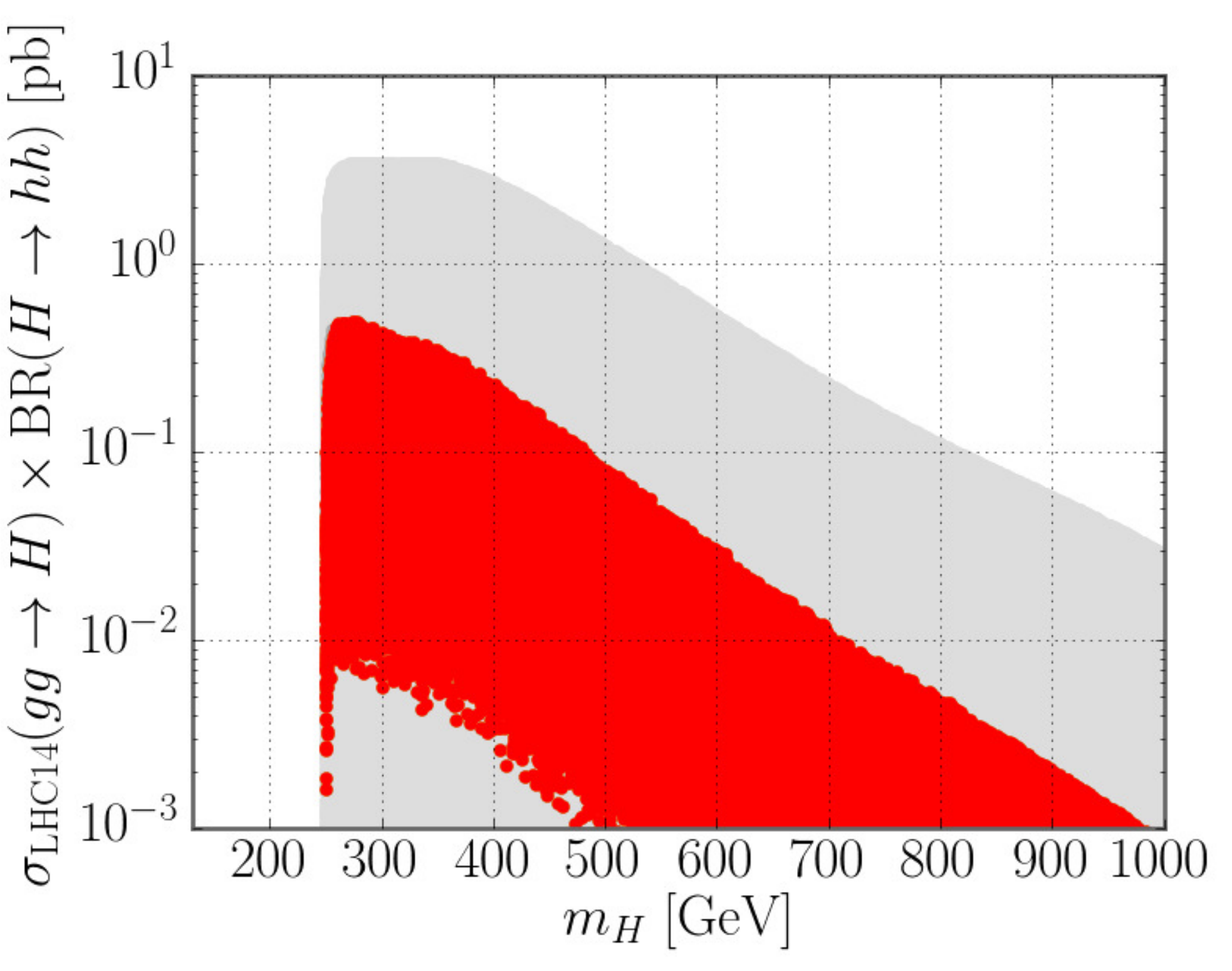}
}
\caption{\label{fig:lhcsig} LHC signal rates of the heavy Higgs boson $H$ decaying into SM particles (\emph{a}) or into two light Higgs bosons, $\Htohh$, (\emph{b}), in dependence of the heavy Higgs mass, $m_H$, for a center-of-mass (CM) energy of $14~\TeV$. Shown are regions which are still allowed after all constraints are taken into account: Red and yellow regions correspond to agreement with the Higgs signal strength measurements at the $1\sigma$ and  $2\sigma$ level, respectively, blue points comply with direct experimental searches but do not agree with the Higgs signal strength within $2\sigma$. Light gray points denote scan points that are excluded by either perturbative unitarity, perturbativity of the couplings, RGE running or the $W$ boson mass, while dark gray points denote regions in parameter space that obey these constraints but are excluded by direct searches.}
\end{figure}

The relatively strong constraints on the mixing angle lead to a {significant} suppression of the direct production {rates} of {the heavy Higgs boson} at LHC run 2. Fig.~\ref{fig:lhcsig} shows the predicted production cross section at $14\,\TeV$ after all constraints have been taken into account. {The} production cross sections rapidly decrease with higher masses $m_H$ {due to} both the stronger constraints on the mixing angle (cf.~Fig.~\ref{fig:sinamw}) {and} a reduction of the available phase space for higher masses. The cross section for direct production {in gluon fusion} and {successive} decay into SM final states ranges from about $10~\pb$ {at lower masses} to about $10~\fb$ for masses around $800~ \GeV$. Note that in order to obtain the predictions for a particular SM decay mode, $H\to XX$, these numbers need to be multiplied by a factor of $\mathrm{BR}(H\to XX)/\mathrm{BR}(H\to \text{SM})$, where $\mathrm{BR}(H\to \text{SM})$ is the sum over all branching ratios of Higgs decays into SM particles according to Eq.~(\ref{eq:brdefs}). Taking into account the current design strategy for the LHC run (cf.~e.g.~Ref.~\cite{lhc_lumi}) and expecting an integrated luminosity of about 100 $\fb^{-1}$ and 300 $\fb^{-1}$ before the shutdowns in 2019 and 2023, respectively, this translates into the fact that at least $\mathcal{O}\lb 10^3\rb$ heavy Higgs bosons could be produced in that mass range in optimistic scenarios. For the $hh$ final state, on the other hand, cross sections are about an order of magnitude lower. {A comparison of current exclusion limits from LHC $\Htohh$ searches with the predictions in the viable parameter space will be given in Section~\ref{sec:bms}.}

Note that these plots were obtained using a {\sl simple rescaling} of production cross section of a SM Higgs boson of the same mass as given in Ref.~\cite{Heinemeyer:2013tqa}, {i.e.~}contributions due to interference with the additional scalar {are not included}. Tools which can handle these have been presented e.g.~in Refs.~\cite{Maina:2015ela,Kauer:2015hia,Ballestrero:2015jca,Dawson:2015haa}. {These studies}, however, focus on effects on the line-shape of the heavy scalar boson {\sl after} a possible discovery. {Moreover, thus far, their} calculations neglect additional higher order corrections, {whereas these have been calculated to great precision for the SM Higgs boson and are included in Fig.~\ref{fig:lhcsig}}~\cite{Heinemeyer:2013tqa}. {For the future, it would be desirable to perform a dedicated study of interference effects including higher order corrections for the benchmark points presented in this work in order to estimate their effects (and the systematic uncertainty introduced here by neglecting them).}

\subsubsection{Low mass region}
In the low mass region, {where the heavier Higgs state takes the role of the discovered Higgs boson,} $m_H\,\sim\,125\,\GeV$, the parameter space is extremely constrained {by the Higgs signal strength and exclusion limits from LEP Higgs searches}~\cite{Alcaraz:2006mx}. The updated experimental results do not change the limits presented in Ref.~\cite{Robens:2015gla}. We review these limits in Tab.~\ref{tab:lowscale}. Note that in {the low mass region the couplings of the heavy Higgs boson at $125\,\GeV$ become SM-like} for $|\sin\al|\,=\,1$. 
\begin{table}
\begin{tabular}{| c | c | c | c | c|}
\toprule
$m_h~[\GeV]$& $|\sin\al|_\text{min, {HB}}$ & $|\sin\al|_\text{min, {HS}}$ &$(\tan\be)_\text{max}$&$(\tan\be)_{\text{no}~\Htohh} $\\
\colrule
{$120$} & {0.410}  & {0.918}    & {8.4}&-- \\
$110$&${0.819}$&${0.932}$&${9.3}$&--\\
$100$&${0.852}$&${0.891}$&$10.1$&--\\
$90$&${0.901}$& -- &$11.2$&-- \\
$80$&${0.974}$&--&$12.6$&--\\
$70$&${0.985} $&--&$14.4$&--\\
$60$&${0.978}$&${0.996}$&$16.8$&{0.21}\\
$50$&${0.981}$&${0.998}$&$20.2$&{0.20}\\
$40$&${0.984}$&${0.998}$&$25.2$&{0.18}\\
{$30$} &{0.988}&{0.998}& {33.6}&{0.16} \\
{$20$} &{0.993}&{0.998}&{50.4}&{0.12} \\
{$10$} &{0.997}&{0.998}&{100.8}&{0.08} \\
\botrule
\end{tabular}
\caption{\label{tab:lowscale} Limits on $\sin\al$ and $\tan\be$ in the low mass scenario for various light Higgs masses $m_h$ and $\tan\be=1$. In the second column we give the lower limit on $\sin\al$ stemming from exclusion limits from LEP or LHC Higgs searches (evaluated with \HB). If the lower limit on $\sin\al$ obtained from the Higgs signal rates (evaluated with \HS) results in stricter limits, they are displayed in the third column. The fourth column displays the upper limit on $\tan\be$ that stems from perturbative unitarity in the complete decoupling case ($|\sin\al|\,=\,1$). In the fifth column we give the $\tan\be$ value for which $\Gamma_{H\rightarrow hh}=0$ is obtained given the maximal mixing angle allowed by the Higgs exclusion limits (second column). At this $\tan\be$ value, the $|\sin\alpha|$ limit obtained from the Higgs signal rates (third column) is abrogated. The table is taken from Ref.~\cite{Robens:2015gla}.}
\end{table}

Tab.~\ref{tab:lowlhc} gives the direct production cross section {in gluon fusion} for the {undiscovered light Higgs state} at a 8 and 14 \TeV~LHC, respectively. Again, the production cross section stems from a simple rescaling of the {corresponding cross section for a SM Higgs boson} of that mass \cite{Heinemeyer:2013tqa,grazzini}.

\begin{table}
\begin{tabular}{| c | c | c |||  c | c | c |}
\toprule
$m_h~[\GeV]$& $\sigma_{gg}^{8\,\TeV}[\pb]$& $\sigma_{gg}^{14\,\TeV}[\pb]$ & $m_h~[\GeV]$& $\sigma_{gg}^{8\,\TeV}[\pb]$& $\sigma_{gg}^{14\,\TeV}[\pb]$ \\
\colrule
{$120$} &3.28 &8.41 &$60$&0.63&1.38\\
$110$&3.24&8.17 &$50$&0.45&0.96\\
$100$&6.12&15.10&$40$&0.76 & 1.59\\
$90$&6.82&16.47 &$30$&1.60  & 3.09\\
$80$&2.33&5.41 &$20$& 5.04 &8.97\\
$70$&1.72&3.91 &$10$& 18.44 & 29.74 \\
\botrule
\end{tabular}
\caption{\label{tab:lowlhc}
Maximally allowed {cross section for light Higgs production in gluon fusion}, $\sigma_{gg}=\lb \cos^2\al \rb_\text{max}\times\sigma_{gg,\text{SM}}$, at the LHC at CM energies of $8$ and $14~\TeV$ after all current constraints have been taken into account, {corresponding to} the mixing angles from Tab.~\ref{tab:lowscale}. This is an updated version of Tab.~V in Ref.~\cite{Robens:2015gla}.}
\end{table}

\clearpage

\subsubsection{Intermediate mass region}

\begin{figure}
\centering
\subfigure[~{$(m_h,m_H)$ plane}.]{
\includegraphics[width=0.48\textwidth]{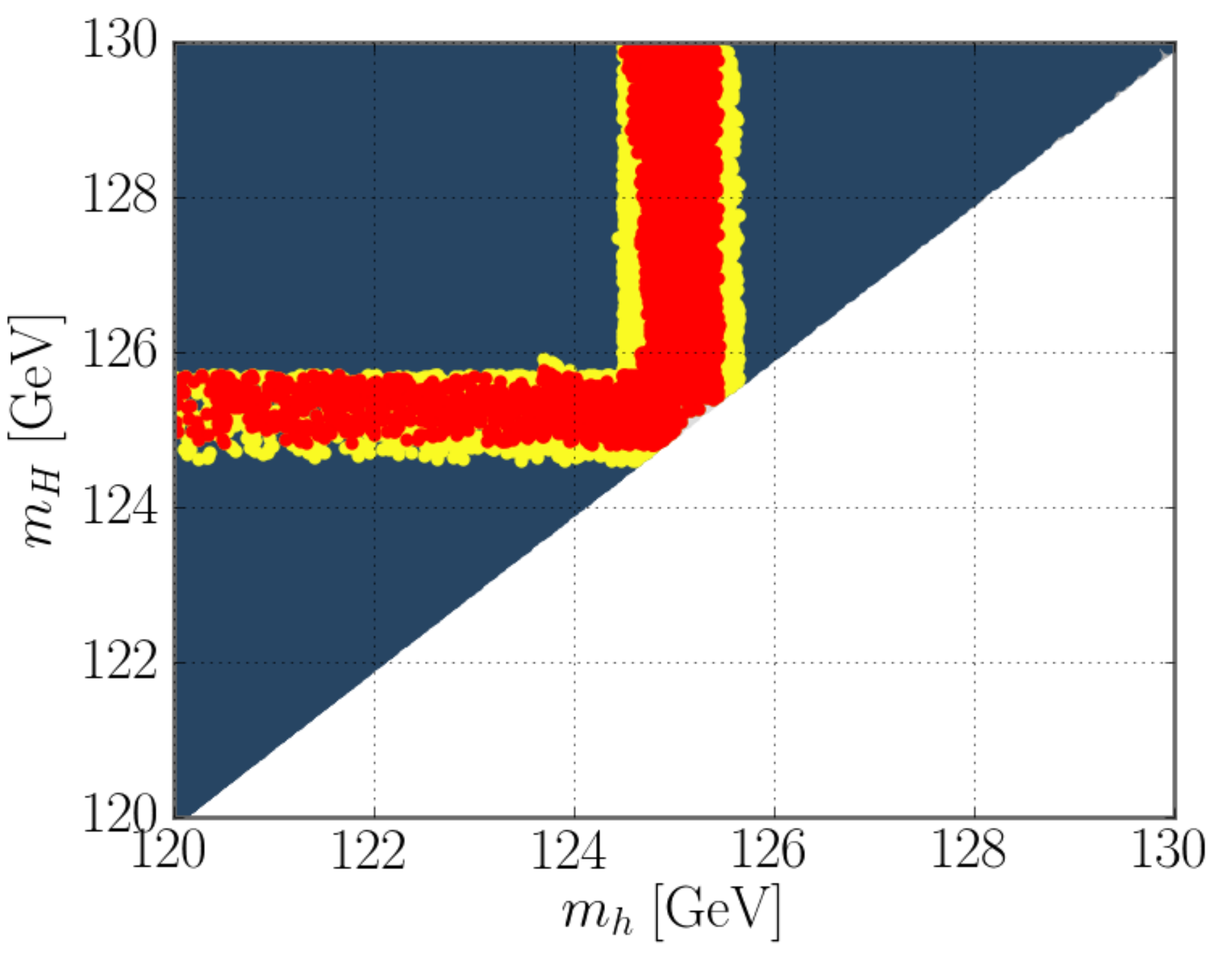}
}
\hfill
\subfigure[~{$(m_h,\sin\al)$ plane}.]{
\includegraphics[width=0.48\textwidth]{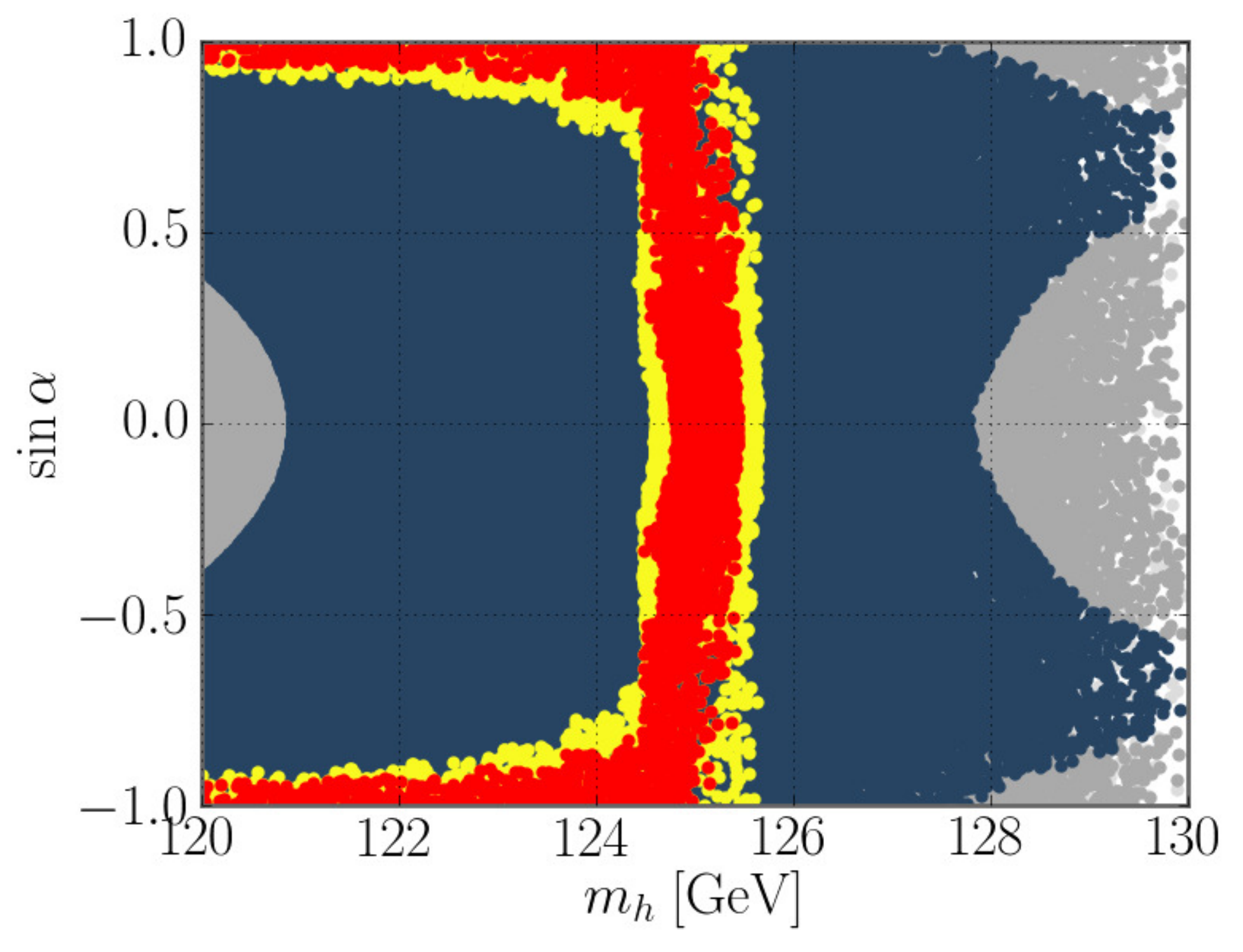}
}
\caption{\label{fig:internew} Parameter space for the intermediate mass region after taking all constraints into account. The color coding follows Fig. \ref{fig:lhcsig}.}
\end{figure}

The intermediate mass region, where both Higgs bosons have masses between $120~\GeV$ and $130~\GeV$, was originally discussed in Ref.~\cite{Robens:2015gla}. In this mass region the observed Higgs signal at $125~\GeV$ may be due to a signal overlap of both Higgs bosons, depending on the mass separation and the mass resolution of the experimental analysis. We show the allowed parameter space in the $(m_h, m_H)$ and $(m_h,\sin\al)$ plane from the updated fit in Fig.~\ref{fig:internew}. The updated signal strength observables in \HSv{1.4.0} yield only marginal improvements in the constrained parameter space, while the updated limits from direct Higgs searches are irrelevant in this mass region.

\section{Benchmark Scenarios for LHC Run 2}
\label{sec:bms}

The benchmark {scenarios} that are presented in this section are chosen such that they feature the {\sl maximally} allowed production cross section at {the} LHC. {We first present the benchmark scenarios for the high mass region, where the light Higgs plays the role of the discovered SM-like Higgs at $125\,\GeV$, and then turn to the low mass range, where the heavy Higgs state is the SM-like Higgs boson.}\footnote{See also Ref.~\cite{Costa:2015llh} for recent benchmark point suggestions within the complex singlet model.}
\subsection{{High mass region}}

 We distinguish between two different search channels:
\begin{itemize}
\item{} {\bf Higgs decays into SM particles:} Maximizing the production cross section corresponds to maximizing the parameter~\cite{Pruna:2013bma}
\begin{\eqn*}
\kappa\,\equiv \,\frac{\sigma}{\sigma_\text{SM}}\times\text{BR} (H\to \mathrm{SM})=\sin^4\al \,\frac{\Gamma_\text{SM,tot}}{\Gamma_\text{tot}}. \label{Eq:kappa}
\end{\eqn*}
In general, following Eq.~\eqref{eq:brdefs}, Higgs decays into SM particles follow the hierarchy of the branching ratios of a SM Higgs of the same mass. This, together with the observation that the branching ratio for $H\,\rightarrow\,hh$ is $\mathcal{O}\lb 0.2 \rb$ in large parts of the parameter space, translates into the fact that for most of the high mass region the dominant decay mode is $H\to WW$.
\item{} {\bf Higgs decays into two light Higgs bosons, $\Htohh$}: Here, the parameter
\begin{\eqn*}
\kappa'\,\equiv\,\frac{\sigma}{\sigma_\text{SM}} \times\text{BR}(\Htohh)=\sin^2\al\,\frac{\Gamma_{H\rightarrow hh}}{\Gamma_\text{tot}},\label{Eq:kappaprime}
\end{\eqn*}
is maximized to obtain the largest possible signal yield.
\end{itemize}

\begin{figure}[t]
\centering
\subfigure[~Heavy Higgs signal rate with SM particles in the final state. {We display the observed and expected $95\%~\text{C.L.}$ limits from the CMS combination of SM Higgs searches~\cite{CMS:aya} as well as from the $H\to VV~(V=W,Z)$ search~\cite{Khachatryan:2015cwa}.}]{
\includegraphics[width=0.48\textwidth]{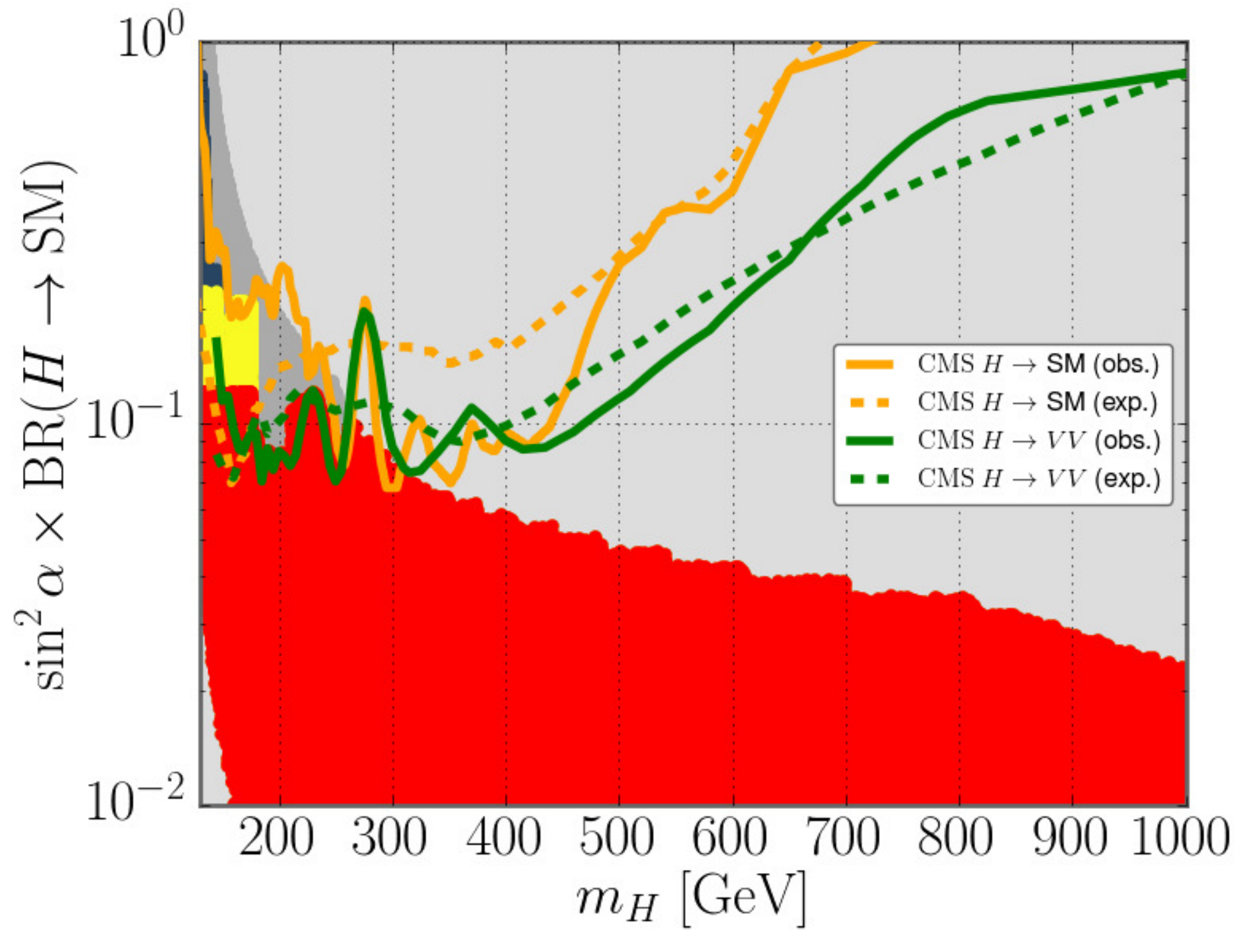}
}
\hfill
\subfigure[~Heavy Higgs signal rate with light Higgs bosons in the final state. We display the current expected and observed $95\%~\text{C.L.}$ limits from {the ATLAS $\Htohh$ search (combination of various final states)~\cite{Aad:2015xja} and} CMS $\Htohh$ searches with $\gamma\gamma b\bar{b}$~\cite{Khachatryan:2016sey} and $b\bar{b}b\bar{b}$~\cite{Khachatryan:2015yea} final states.]{
\includegraphics[width=0.48\textwidth]{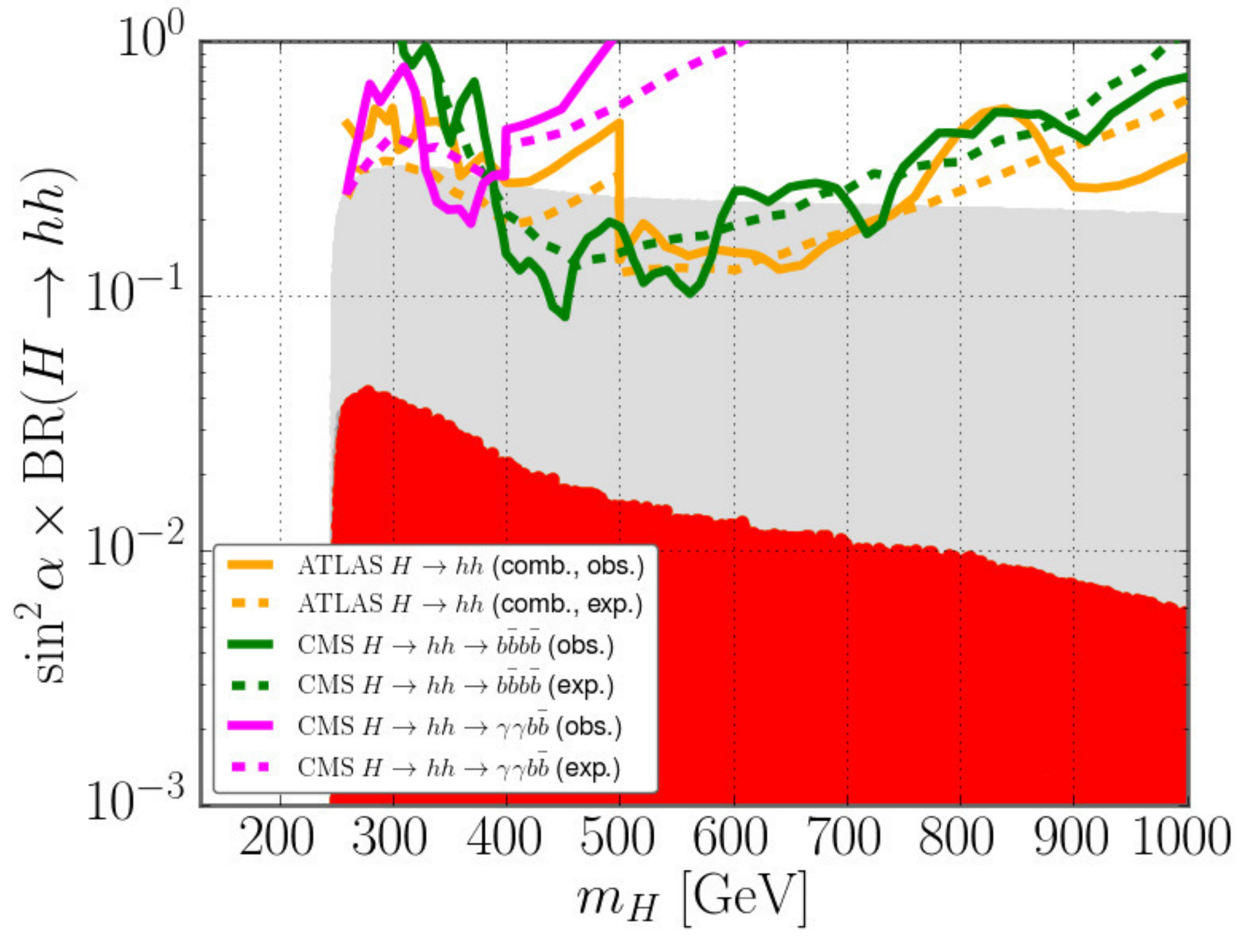}
}
\caption{\label{fig:lhcsig_abs} Collider signal rates of the heavy Higgs boson $H$ decaying into SM particles (\emph{a}) or into two light Higgs bosons, $\Htohh$, (\emph{b}), in dependence of the heavy Higgs mass, $m_H$. {The color coding is the same as in Fig.~\ref{fig:lhcsig}.} The rates are normalized to the inclusive SM Higgs production cross section at the corresponding mass value~\cite{Dittmaier:2011ti,Dittmaier:2012vm,Heinemeyer:2013tqa}.}
\end{figure}

Figure \ref{fig:lhcsig_abs} shows the {allowed} range of these two quantities, after all constraints have been taken into account. For {the Higgs decay channel into SM particles}, we see that searches from CMS pose important constraints for $m_H\,\lesssim\,{400}\,\GeV$. For the {Higgs-to-Higgs} decay channel $\Htohh$, on the other hand, both ATLAS~\cite{Aad:2015xja} and CMS~\cite{Khachatryan:2015yea,Khachatryan:2016sey} searches are not yet sensitive enough to {exclude points that are not already in conflict with other constraints.} 

{We} quantify {the} benchmark {scenarios} {for both signal channels} in this regime {by} consider{ing} the {\sl maximally} allowed mixing angle together with the {\sl maximal} and {\sl minimal} branching ratio {for the decay $\Htohh$, respectively}. While these maximal and minimal points define {\sl benchmark points}, all {$\text{BR}(\Htohh)$} values in between are in principle allowed. Therefore, an {interpolation} between the {minimal and} maximal {values} defines a {higher-dimensional benchmark scenario ({\sl benchmark slope} or {\sl plane})}, where the additional third parameter (cf.~Eq.~\eqref{eq:par_choices1}-\eqref{eq:par_choices4}) is floating. 

We furthermore distinguish scenarios for which the {$\Htohh$} on-shell decay mode is {kinematically} allowed {or} forbidden. As we neglect all other triple and quartic Higgs self-couplings apart from $\mu'$, and work in the on-shell {approximation}, $\tan\be$ only influences the collider phenomenology for regions in parameter space where the decay {$\Htohh$} is {kinematically} allowed, {i.e.~for heavy Higgs masses $m_H \ge 2m_h \approx 250\,\GeV$}. {For lower masses $\tan\beta$} is irrelevant for the phenomenology considered here. However, {to be consistent,} we recommend to still keep the values within the respective parameter regions allowed by perturbativity and perturbative unitarity.

 Benchmark {scenarios} for both cases are given in Tab.~\ref{tab:highm1} and \ref{tab:highm2}, respectively.  Parameter ranges which are not explicitly listed can to a first approximation be linearly interpolated.
\begin{table}
\begin{center}
\begin{tabular}{|c|c|c|||c|c|c|}
\toprule
$m_H [\GeV]$&$|\sin\alpha|_\text{max}$&$\tan\beta_\text{max}$ &$m_H [\GeV]$&$|\sin\alpha|_\text{max}$&$\tan\beta_\text{max}$\\ 
\colrule
130&	0.42&1.79 &195&0.28&1.22\\
135&	0.38&1.73 &200&0.29&1.19\\
140&	0.36&1.69&210&0.28&1.14\\
145&	0.35&1.62&215&0.33&1.12\\
150&	0.34&1.57 &220&0.34&1.10\\
160& 0.36&1.49&230&0.35&1.05\\
180&0.30&1.32&235&0.34&1.03\\
185&0.27&1.28&240&0.31&1.00\\
190&0.29&1.26&245&0.28&0.98\\
\botrule
\end{tabular}
\end{center}
\caption{\label{tab:highm1} Benchmark points for mass ranges where the onshell decay $\Htohh$ is kinematically forbidden. Maximal values of $\tan\be$ were calculated at the maximal mixing angle, and should be applied for consistency reasons.}
\end{table}

\begin{table}
\begin{tabular}{|c|c|c|c|||c|c|c|c|}
\toprule
$m_H [\GeV]$&$|\sin\al|_\text{max}$&$BR^{\Htohh}_\text{min}$&$BR^{\Htohh}_\text{max}$ & $m_H [\GeV]$&$|\sin\al|_\text{max}$&$BR^{\Htohh}_\text{min}$&$BR^{\Htohh}_\text{max}$ \\
\colrule
255 &0.31	& 0.09 &0.27 &430&	0.25&0.19&0.30 \\
260&	0.34 & 0.11 & 0.33 &470&	0.24&0.19 & 0.28 \\
265&	0.33&0.13  &0.36 &520&	0.23&0.19 & 0.26 \\
280&	0.32& 0.17&0.40 &590&	0.22& 0.19 &0.25 \\
290&	0.31&0.18&0.40 &665&	0.21& 0.19& {0.24} \\
305&	0.30&0.20&0.40 &770&	0.20& 0.19&0.23 \\
325&	0.29& 0.21 & 0.40 &875&	0.19&0.19&0.22\\
345&	0.28&0.22&0.39 &920&0.18&0.19&0.22\\
365&	0.27& 0.21 & 0.36 &975&0.17&0.19&0.21\\
395&	0.26& 0.20 &0.32 &1000&0.17&0.19&0.21\\
\botrule
\end{tabular}
\caption{\label{tab:highm2} Maximal and minimal allowed branching ratios of the decay $\Htohh$, taken at the maximally allowed value of $|\sin\al|$. Note that mininal values for the $\text{BR}(\Htohh)$ stem from $\sin\al\,\leq\,0$.}
\end{table}

In addition, we also list exemplary {\sl benchmark points} for this mass region in Tables \ref{tab:BMH2} and \ref{tab:BHM1}, where we additionally give {the predictions for} other {relevant} decay modes. When{ever} kinematically accessible, we provide two benchmark points {for every heavy Higgs mass, representing the} maximal and minimal branching ratio for the $\Htohh$ decay, respectively.\footnote{Electroweak corrections to the decay $\Htohh$ have been presented for some of these benchmark points in Ref.~\cite{Bojarski:2015kra}.} The mixing angle is always {chosen such that the production rate of the additional scalar is maximized.}
\begin{table}[h]
\begin{center}
\begin{tabular}{|l|l|}
\toprule  
\multicolumn{2}{|c|}  {\bf Benchmark Scenarios for the Real Singlet}\cr
\colrule
Main Features & real singlet extension, with two vevs and no hidden sector interaction \cr
&  with heavy Higgs $H$ and light Higgs $h$. \cr
\colrule
Fixed parameters	& $M_h$ = 125.1 GeV or $M_H$ = 125.1 GeV. \cr
Irrelevant parameters & $\tan\beta$ when{ever} channel $\Htohh$ {kinematically} not accessible.\cr
{additional comments} & predictions at LO, factorized production and decay; \cr
 & \emph{a},\emph{b} signify maximal and minimal $\text{BR}(\Htohh)$; for \emph{b}, $\sin\alpha\,<\,0$;\cr
& any values  for $\tan\beta$ between scenario \emph{a} and \emph{b} are allowed. \cr 
\toprule
\multicolumn{2}{|c|}  {Production cross sections at 14 TeV [pb]  and branching fractions}\cr
\colrule
\multicolumn{2}{|c|}  {BHM300 \emph{a},\emph{b} \hspace*{\fill}}\cr
\colrule
Spectrum & $M_H$=300 GeV, $|\sin\alpha|= {0.31},\,\tan\beta~(a)= {0.79}, \,\tan\beta~(b)= {0.79} $ \cr
$\sigma(gg \rightarrow h)$ &  {44.91} \cr
$\sigma(gg \rightarrow H)$ &  {1.09} \cr
BR($\Htohh$) & $0.41~(a)$, $0.17~(b)$ \cr
BR($H \rightarrow W W$) & $0.41~(a)$, $0.57~(b)$ \cr
BR($H \rightarrow Z Z$) &  $0.18~(a)$, $0.25~(b)$ \cr
\colrule
\multicolumn{2}{|c|}  {BHM400 \emph{a},\emph{b} \hspace*{\fill}}\cr
\colrule
Spectrum & $M_H$=400 GeV, $|\sin\alpha|=0.26,\,\tan\beta~(a)=0.58,\,\tan\beta~(b)=0.59$ \cr
$\sigma(gg \rightarrow h)$ &   {46.32} \cr
$\sigma(gg \rightarrow H)$ &   {0.76} \cr
BR($\Htohh$) & $0.32~(a)$, $0.20~(b)$\cr
BR($H \rightarrow W W$) &  $0.40~(a)$, $0.47~(b)$ \cr
BR($H \rightarrow Z Z$) & $0.18~(a)$, $0.22~(b)$ \cr
BR($H \rightarrow t \bar{t}$) & $0.10~(a)$, $0.12~(b)$ \cr
\colrule
\multicolumn{2}{|c|}  {BHM500 \emph{a},\emph{b} \hspace*{\fill}}\cr
\colrule
Spectrum & $M_H$=500 GeV, $|\sin\alpha|=0.24,\,\tan\beta~(a)=0.44,\,\tan\beta~(b)=0.46$ \cr
$\sigma(gg \rightarrow h)$ &  {46.82}\cr
$\sigma(gg \rightarrow H)$ &  {0.31} \cr
BR($\Htohh$) & $0.26~(a)$, $ {0.19}~(b)$ \cr
BR($H \rightarrow W W$) &  $0.41~(a)$, $ {0.44}~(b)$ \cr
BR($H \rightarrow Z Z$) &  $0.19~(a)$, $ {0.21}~(b)$ \cr
BR($H \rightarrow t \bar{t}$) & $0.14~(a)$, $ {0.16}~(b)$ \cr
\botrule
\end{tabular}
\end{center}
\caption{\label{tab:BMH2} Benchmark scenarios for the high mass region for fixed masses and $|\sin\alpha|$, floating $\tan\beta$ (between scenarios \emph{a} and \emph{b}).  {Reference production cross sections have been taken from the upcoming CERN Yellow Report 4 by the LHC Higgs Cross Section Working Group~\cite{YR4}.}}
\end{table}
\newpage
\begin{table}
\begin{center}
\begin{tabular}{|l|l|}
\toprule
\multicolumn{2}{|c|}  {\qquad Production cross sections at 14 TeV [pb]  and branching fractions (\emph{continued}) \qquad}\cr
\colrule
\multicolumn{2}{|c|}  {BHM600 \emph{a},\emph{b} \hspace*{\fill}}\cr
\colrule
Spectrum & $M_H$=600 GeV, $|\sin\alpha|=0.22,\,\tan\beta~(a) =0.37,\,\tan\beta~(b)=0.38 $ \cr
$\sigma(gg \rightarrow h)$ &   {47.28} \cr
$\sigma(gg \rightarrow H)$ &  {0.12} \cr
BR($\Htohh$) & $0.25~(a)$, $0.19~(b)$ \cr
BR($H \rightarrow W W$) & $0.41~(a)$, $0.45~(b)$ \cr
BR($H \rightarrow Z Z$) &  $0.21~(a)$, $0.22~(b)$\cr
BR($H \rightarrow t \bar{t}$) & $0.13~(a)$, $0.14~(b)$ \cr
\colrule
\multicolumn{2}{|c|}  {BHM700 \emph{a},\emph{b} \hspace*{\fill}}\cr
\colrule
Spectrum & $M_H$=700 GeV, $|\sin\alpha|=0.21,\,\tan\beta~(a)=0.31,\,\tan\beta~(b)=0.32$ \cr
$\sigma(gg \rightarrow h)$ &  {47.49}\cr
$\sigma(gg \rightarrow H)$ &  {0.050}\cr
BR($\Htohh$) & $0.24~(a)$, $0.19~(b)$ \cr
BR($H \rightarrow W W$) & $0.44~(a)$, $0.47~(b)$ \cr
BR($H \rightarrow Z Z$) & $0.22~(a)$, $0.23~(b)$\cr
BR($H \rightarrow t \bar{t}$) & $0.10~(a)$, $0.11~(b)$ \cr
\colrule
\multicolumn{2}{|c|}  {BHM800 \emph{a},\emph{b} \hspace*{\fill}}\cr
\colrule
Spectrum & $M_H$=800 GeV, $|\sin\alpha|=0.2,\,\tan\beta~(a)=0.25, \tan\beta~(b)=0.27 $ \cr
$\sigma(gg \rightarrow h)$ &  {47.69} \cr
$\sigma(gg \rightarrow H)$ &  {0.022} \cr
BR($\Htohh$) & $0.23~(a)$, $0.19~(b)$ \cr
BR($H \rightarrow W W$) & $0.46~(a)$, $0.48~(b)$ \cr
BR($H \rightarrow Z Z$) &  $0.23~(a)$, $0.24~(b)$ \cr
BR($H \rightarrow t \bar{t}$) & $0.08~(a)$, $0.09~(b)$ \cr
\colrule
\multicolumn{2}{|c|}  {BHM200 \hspace*{\fill}}\cr
\colrule
Spectrum & $M_H$=200 GeV, $|\sin\alpha|=0.29,\,\tan\beta = 1.19 $ \cr
$\sigma(gg \rightarrow h)$ &  {45.50} \cr
$\sigma(gg \rightarrow H)$ &  {1.74} \cr
BR($H \rightarrow \text{SM}$) & as {for} a SM Higgs boson with mass of $200\,\GeV$ \cr
\botrule
\end{tabular}
\end{center}
\caption{\label{tab:BHM1} Benchmark scenarios for the high mass region for fixed masses and $|\sin\alpha|$, floating $\tan\beta$ (between scenarios \emph{a} and \emph{b}).  {Reference production cross sections have been taken from the upcoming CERN Yellow Report 4 by the LHC Higgs Cross Section Working Group~\cite{YR4}.}}
\end{table}

\clearpage
\subsection{{Low mass region}}
For the case that {the heavier Higgs boson is taken to be the discovered SM-like Higgs boson with} $m_H\,\sim\,125\,\GeV$, $|\sin\al|\,=\,1$ corresponds to the SM limit, and deviations from this value parametrize the new physics contributions. As in the high mass region, the following  channels are interesting:
\begin{itemize}
\item{}{\sl Direct production} of the lighter {Higgs} state {$h$} {and successive decay into SM particles,}
\item{}{\sl Decay} {of the SM-like Higgs boson $H$ into the lighter Higgs states,} $\Htohh$.
\end{itemize}
For the direct production {of the light Higgs state} smaller $|\sin\al|$ values are of interest, {as the cross section scales with $\cos^2\al$}. We provide the minimally allowed values for $|\sin\al|$ in Tab.~\ref{tab:lowscale}. Tab.~\ref{tab:lowlhc} lists the respective direct production cross sections at 8 and 14 \TeV. These values can directly be used as benchmark scenarios for collider searches for direct light Higgs production.

For the second channel --- the decay of the SM-like Higgs into two lighter Higgs states --- we list maximal branching ratios for the decay $\Htohh$ in Tab.~\ref{tab:maxbrlight}. {As long as the decay $\Htohh$ is kinematically accessible, the maximal value of its branching ratio, $\text{BR}(\Htohh) \simeq 0.259$, is not dependent on the light Higgs mass.} The lighter Higgs bosons then decay further according to the branching ratios of a SM Higgs of the respective mass. {A first experimental search of this signature with the light Higgs boson decaying into $\tau$ lepton pairs in the mass range $m_h \in [5, 15]\,\GeV$ has already been performed by the CMS experiment~\cite{CMS:2015iga}.}

\begin{table}[b]
\begin{tabular}{|c|c|c|c}
\toprule
$m_h [\GeV]$&$\sin\al$&$BR^{\Htohh}_\text{max}$ \\ 
\colrule
60&0.9996 &0.259\\
50&0.9999&0.259\\
40&0.9999&0.259\\
30&0.9999&0.259\\
20&0.9998&0.259\\
10&0.9999&0.259\\
\botrule
\end{tabular}
\caption{\label{tab:maxbrlight}Maximal branching ratios for $\Htohh$. This BR can always be zero for the choice $\tan\be\,=\,-\cot\,\al$. }
\end{table}
We present benchmark {points} for fixed masses in Tab.~\ref{tab:BHL}. Here, $|\sin\al|$ values closer to unity are needed in order to obtain maximal branching ratios for this channel, which in turn leads to the reduction of direct production for the lighter state by almost an order of magnitude with respect to the values presented in Tab.~\ref{tab:lowlhc}. {Again, we recommend to scan over $\tan\beta$ between the values of scenario \emph{a} and \emph{b} (thus defining a higher dimensional benchmark scenario) in order to obtain a range of possible branching ratios.}

\begin{table}
\begin{center}
\begin{tabular}{|l|l|}
\toprule
\multicolumn{2}{|c|}  {BHM60 \emph{a},\emph{b} \hspace*{\fill}}\cr
\colrule
Spectrum & $M_h$=60 GeV, $|\sin\alpha|=0.9997,\,\tan\beta~(a) =3.48,\,\tan\beta~(b)=0.025 $ \cr
$\sigma(gg \rightarrow h)$ &  0.10 \cr
$\sigma(gg \rightarrow H)$ &  {49.65} \cr
BR($\Htohh$) & $0.26~(a)$, $0~(b)$ \cr
BR($H \rightarrow \text{SM}$) & rescaled by $0.74~(a)$, as in SM $(b)$ \cr
\colrule
\multicolumn{2}{|c|}  {BHM50 \emph{a},\emph{b} \hspace*{\fill}}\cr
\colrule
Spectrum & $M_h$=50 GeV, $|\sin\alpha|=0.9998,\,\tan\beta~(a)=3.25,\,\tan\beta~(b)=0.020$ \cr
$\sigma(gg \rightarrow h)$ & {0.098}\cr
BR($\Htohh$) & $0.26~(a)$, $0~(b)$ \cr
BR($H \rightarrow \text{SM}$) & rescaled by $0.74~(a)$, as in SM $(b)$ \cr
\colrule
\multicolumn{2}{|c|}  {BHM40 \emph{a},\emph{b} \hspace*{\fill}}\cr
\colrule
Spectrum & $M_h$=40 GeV, $|\sin\alpha|=0.9998,\,\tan\beta~(a)=3.13, \tan\beta~(b)=0.020 $ \cr
$\sigma(gg \rightarrow h)$ & 0.16 \cr
BR($\Htohh$) & $0.26~(a)$, $0~(b)$ \cr
BR($H \rightarrow \text{SM}$) & rescaled by $0.74~(a)$, as in SM $(b)$ \cr\hline
\multicolumn{2}{|c|}  {BHM30 \emph{a},\emph{b} \hspace*{\fill}}\cr
\colrule
Spectrum & $M_h$=30 GeV, $|\sin\alpha|=0.9998,\,\tan\beta~(a)=3.16, \tan\beta~(b)=0.020 $ \cr
\colrule
$\sigma(gg \rightarrow h)$ & 0.31 \cr
BR($\Htohh$) & $0.26~(a)$, $0~(b)$ \cr
BR($H \rightarrow \text{SM}$) & rescaled by $0.74~(a)$, as in SM $(b)$ \cr
\colrule
\multicolumn{2}{|c|}  {BHM20 \emph{a},\emph{b} \hspace*{\fill}}\cr
\colrule
Spectrum & $M_h$=20 GeV, $|\sin\alpha|=0.9998,\,\tan\beta~(a)=3.23, \tan\beta~(b)=0.020 $ \cr
$\sigma(gg \rightarrow h)$ & 0.90 \cr
BR($\Htohh$) & $0.26~(a)$, $0~(b)$ \cr
BR($H \rightarrow \text{SM}$) & rescaled by $0.74~(a)$, as in SM $(b)$ \cr
\colrule
\multicolumn{2}{|c|}  {BHM10 \emph{a},\emph{b} \hspace*{\fill}}\cr
\colrule
Spectrum & $M_h$=10 GeV, $|\sin\alpha|=0.9998,\,\tan\beta~(a)=3.29, \tan\beta~(b)=0.020 $ \cr
$\sigma(gg \rightarrow h)$ & 2.98 \cr
BR($\Htohh$) & $0.26~(a)$, $0~(b)$ \cr
BR($H \rightarrow \text{SM}$) & rescaled by $0.74~(a)$, as in SM $(b)$ \cr
\colrule
\end{tabular}
\end{center}
\caption{\label{tab:BHL} {Low mass} benchmark {scenarios} {for the Higgs-to-Higgs decay signature} for fixed masses and $|\sin\alpha|$, floating $\tan\beta$ (between scenarios \emph{a} and \emph{b}). In scenario \emph{b} we have $\tan\beta\,=\,-\cot\,\alpha$. {The $|\sin\alpha|$} values have been optimized for scenario \emph{a}, which in turn leads to a suppression of direct production for the lighter state. For direct production of the lighter scalar, {the parameters} in Tab.~\ref{tab:lowscale} and \ref{tab:lowlhc} should be used. For BHM50 - BHM10, the production cross section for the SM like Higgs is $\sigma(gg\,\rightarrow\,H)= {49.66}\, \pb$.  {Reference production cross sections have been taken from the upcoming CERN Yellow Report 4 by the LHC Higgs Cross Section Working Group~\cite{YR4}.}}
\end{table}

\section{Conclusions}
\label{sec:conclude}

In this {paper} we have {revisited and updated the constraints} on the parameter space of {the real scalar singlet extension of} the SM. In comparison with the previous results presented in Ref.~\cite{Robens:2015gla}, {the most important improvements have been made in the constraints from new results in LHC searches for a heavy Higgs boson decaying into vector boson final states, as well as from the ATLAS and CMS combination of the signal strength of the discovered Higgs state.} We found that these modify our previous findings in the mass range $130\,\GeV \leq m_H\,\leq\,250\,\GeV$, where now the direct Higgs searches as well as the {ATLAS and CMS} signal strength combination render the {strongest} constraints on the parameter space. 

Based on these updated results, we have {provided} benchmark {scenarios} for {both the high mass and low mass region for upcoming LHC searches.} {Hereby, we pursued the philosophy of selecting those points which feature a maximal discovery potential in a dedicated collider search of the corresponding signature.} We provided predictions of production cross sections {for} the LHC at 14 \TeV{, and supplemented these with information about the branching fractions of the relevant decay modes}. 
We {encourage} the {experimental collaborations} to {make use} of the{se} benchmark {scenarios in} the current {and upcoming} LHC run{s}.
\section*{Acknowledgements}
{We} thank S. Dawson, C. Englert, M. Gouzevitch, S. Heinemeyer, I. Lewis, A. Nikitenko, M. Sampaio, R. Santos, M. Slawinska, and D. St\"ockinger for useful discussions, as well as G. Chalons, D. Lopez-Val, and G.M. Pruna for fruitful collaboration on earlier related work. {TS is supported in parts by the U.S.~Department of Energy grant number DE-SC0010107 and a Feodor-Lynen research fellowship sponsored by the Alexander von Humboldt foundation.}

\bibliography{main}

\end{document}